\documentclass[12pt,english,aps]{revtex4}
\usepackage[T1]{fontenc}
\usepackage[latin9]{inputenc}
\usepackage[a4paper]{geometry}
\usepackage{amsmath}
\usepackage{graphicx}
\usepackage{amssymb}
\usepackage{esint}
\usepackage{ngerman}
\usepackage{textcomp}


\usepackage{babel}

\begin{document}

\title{Bose-Einstein condensation of photons}

\author{Jan Klaers and Martin Weitz}

\address{Institut f\"ur Angewandte Physik, Universit\"at Bonn, Wegelerstr.
8, 53115 Bonn, Germany}

\pagenumbering{arabic}

\begin{abstract}
We review recent work on the Bose-Einstein condensation of photons
in a dye microcavity environment. Other than for material particles,
as e.g. cold atomic Bose gases, photons usually do not condense at
low temperatures. For Planck\textquoteright{}s blackbody radiation,
the most ubiquitous Bose gas, photon number and temperature are not
independently tunable and at low temperatures the photons simply disappear
in the system\textquoteright{}s walls, instead of massively occupying
the cavity ground mode. In the here described approach, this obstacle
is overcome by a fluorescence-induced thermalization mechanism in
a dye-filled microcavity. Experimentally, both the thermalization
of the photon gas and, at high photon densities, Bose-Einstein condensation
has been observed. This article describes the thermalization mechanism
of the photon gas in detail and summarizes so far performed experimental
work. 
\end{abstract}

\maketitle

\tableofcontents{}

\section{Introduction}

When a gas of particles with given density is cooled to such low temperatures
that the associated de Broglie wavepackets spatially overlap, quantum
statistical effects come into play. Specifically, for a gas of particles
with integer spin (bosons), Bose-Einstein condensation into the ground
state sets in above a critical phase space density. For dilute atomic
gases, this effect has first been observed in 1995 by means of laser
and subsequent evaporative cooling of alkali atoms \cite{Anderson:1995,Davis:1995p2059,Bradley:1997},
see also the article of F. Chevy and J. Dalibard in this volume. The
signature of Bose-Einstein condensation has also been observed for
several solid-state quasi-particles, as exciton-polaritons and magnons,
see e.g. the contributions by Y. Yamamoto and by V. Demokritov and A. Slavin, respectively, in this volume. 

Photons, the quantized particles of light also are bosons, but usually
show no Bose-Einstein condensation. In a blackbody radiator, the chemical
potential of photons vanishes, i.e. the average particle number does
not follow a given conservation law, but adjusts itself to the available
thermal energy \cite{Huang:StatisticalMechanics1987}. This is the
essence of the Stefan-Boltzmann law, linking the total radiation energy
$U$ to the fourth power of the temperature, $U\propto T^{4}$. At
low temperatures, the photon number simply decreases and no macroscopic
occupation of the cavity ground state occurs. Thus, a necessary precondition
for a Bose-Einstein condensation of photons is to find a thermalization
process that allows for an independent adjustment of both photon number
and temperature. The lack of such a mechanism has long prevented the
realization of light sources that are capable of generating single
mode light, without the necessity to be driven out of thermal equilibrium,
as e.g. for a laser. Note that in a laser both the state of the light
field and that of the active medium are far removed from thermal equilibrium
\cite{Siegman:1986}. To some extent, lasing is even a prime example
for a non-equilibrium process, as only the absence of thermal equilibrium
allows for inversion and optical gain. Early theoretical work has
proposed to reach Bose-Einstein condensation of photons by Compton
scattering off a thermal electron gas \cite{Zeldovich:1969p1287}.
Later, Chiao et al. proposed a two-dimensional photon quantum fluid
in a nonlinear Fabry-Perot resonator, where thermalization was sought
from photon-photon scattering \cite{Chiao:PhysicalReviewA1999,Chiao:OpticsCommunications2000,Bolda:2001p514}.
This concept is similar to atom-atom scattering processes in atomic
physics BEC experiments, though the limited non-linearity has so far
prevented a thermalization of the photon gas \cite{Mitchell:2000p1910}.
In other work, the demonstration of (quasi-)equilibrium Bose-Einstein
condensation of exciton-polaritons, mixed states of matter and light,
has been reported \cite{Deng:RevModPhys2010,Kasprzak:2006p443,Balili:2007p1342}.
Here interparticle collisions of the excitons, i.e. the material parts
of the polaritons, act as a thermalization mechanism. In other experiments,
superfluidity of polaritons has been observed \cite{Amo:NaturePhysics2009,Lagoudakis:NaturePhysics2008}. 

\begin{figure}
\begin{centering}
\vspace{-7mm}

\par\end{centering}

\begin{centering}
\includegraphics[width=0.9\columnwidth]{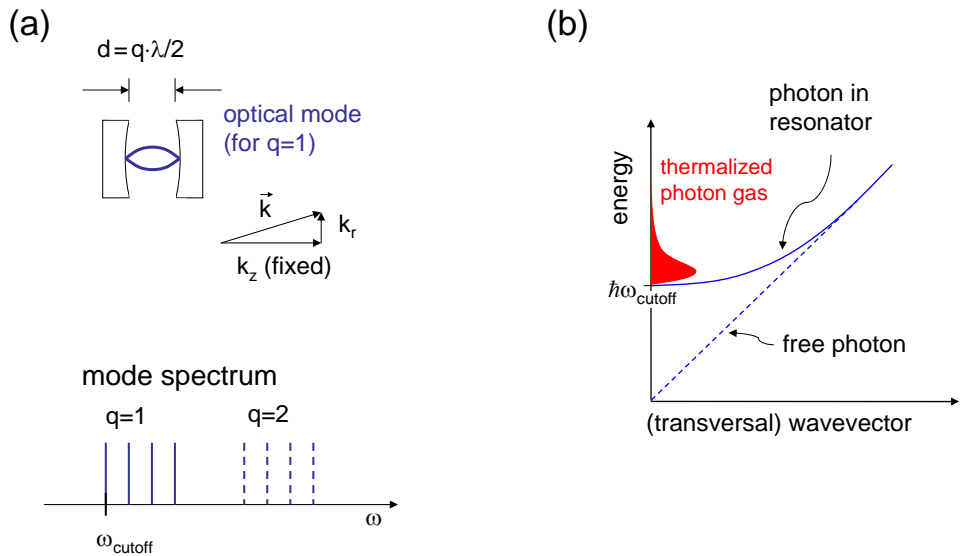}
\par\end{centering}

\begin{centering}
\vspace{-17mm}
\caption{\label{fig:Scheme}(a) Scheme of optical resonator (top) and the cavity
modes (bottom) for the case of a spacing between resonator mirrors
of half an optical wavelength of the lowest optical mode. The resonator
is filled with dye solution. The photon gas thermalizes to the temperature
of the dye solution. (b) Photon dispersion in resonator (solid line)
and the dispersion of a free photon (dashed line). }

\par\end{centering}

\centering{}\vspace{3mm}

\end{figure}

In recent experiments of our group, photon Bose-Einstein condensation
is achieved in a dye-solution filled optical microresonator \cite{Klaers:2010,Klaers:2010p2137,Klaers:2011}.
Thermalization of the photon gas with the dye is achieved by repeated
absorption emission cycles. For such systems it is known that frequent
collisions ($\sim10\,\text{fs}$ timescale) between solvent and dye
molecules causes rapid transverse decoherence at room temperature,
so that the condition of strong light-matter coupling is not met \cite{Yokoyama:1989p2104,DeAngelis:2000p484}.
The distance between the two spherically curved resonator mirrors
is in the micrometer regime, which causes a large frequency spacing
between the longitudinal resonator modes. The latter is of order of
the emission width of the dye molecules. In combination with an intracavity
modification of the spontaneous emission, preferring the emission
to small volume modes (low transversal excitation), a regime is reached,
where to good approximation the resonator is populated only by photons
of a single longitudinal mode number, see Fig. \ref{fig:Scheme}a.
The longitudinal modal quantum number is frozen out and the photon
gas effectively becomes two-dimensional. As is indicated in Fig. \ref{fig:Scheme}b,
the photon dispersion relation acquires quadratic, i.e. particle-like
character, with the frequency of the transverse $\text{TEM}_{00}$
mode acting as a low-frequency cutoff frequency. Furthermore, a harmonic
trapping potential for the photon gas is induced by the mirrors curvature.
Thermal equilibrium of the photon gas with its environment (at room
temperature) is achieved as the photons are absorbed and emitted by
the dye molecules repeatedly. The photon frequencies will accumulate
within a spectral range of order $\sim k_{\text{B}}T/\hbar$ above
the low frequency cutoff. Other than in a blackbody radiator, the
thermalization process allows for an independent adjustment of temperature
and photon number. This becomes clear by noting that the energy of
fluorescence photons, which takes values above the cutoff frequency
of the resonator in energy units ($\hbar\omega_{\text{cutoff}}\simeq2.1\,\text{eV}$),
is far above thermal energy, $k_{\text{B}}T\simeq1/40\,\textrm{eV}$.
Thus, purely thermal excitation of (optical) photons is negligible.
Instead, the number of photons will be determined by the strength
of the optical pumping. One can show that the photon gas confined
in the resonator is formally equivalent to a harmonically trapped
two-dimensional gas of massive bosons with effective mass $m_{\text{eff}}=\hbar\omega_{\text{cutoff}}n_{0}^{2}/c^{2}$,
where $n_{0}$ denotes the refractive index of the medium and $c$
the vacuum speed of light. For such a system it is well known that
a Bose-Einstein condensate exists at a finite temperature \cite{Bagnato:1991p552,Mullin:1997p1821}.
In recent experiments, we have both observed thermalization of the
photon gas in the dye-filled microcavity system \cite{Klaers:2010p2137}
as well as Bose-Einstein condensation \cite{Klaers:2010}. The so
far observed properties of the photon Bose-Einstein condensates in
many respects resemble that of atomic gases, while the approximately
10 orders of magnitude smaller effective photon mass allows for transition
temperatures in the room temperature regime. 

Bose-Einstein condensation and superfluidity are two closely related
phenomena \cite{Griffin:1995}. While the former is connected to equilibrium
properties, the latter deals with transport properties. Bose-Einstein
condensation is in principle possible with an ideal gas, while the
presence of superfluidity requires interparticle interactions. In
future, it remains to be experimentally verified whether the photon
Bose-Einstein condensate also exhibits superfluidity. 

In the following, section II describes the fluorescence induced thermalization
mechanism of the photon gas and section III the statistical theory of
the trapped photon gas. Further, section IV reviews experiments on
the thermalization process and section V corresponding results on
Bose-Einstein condensation. Finally, section VI gives conclusions.

\section{Fluorescence induced thermalization}

\subsection{\label{sub:Kennard-Stepanov}Kennard-Stepanov theory of dye spectra}

\begin{figure}
\begin{centering}
\vspace{0mm}

\par\end{centering}

\begin{centering}
\includegraphics[width=0.8\columnwidth]{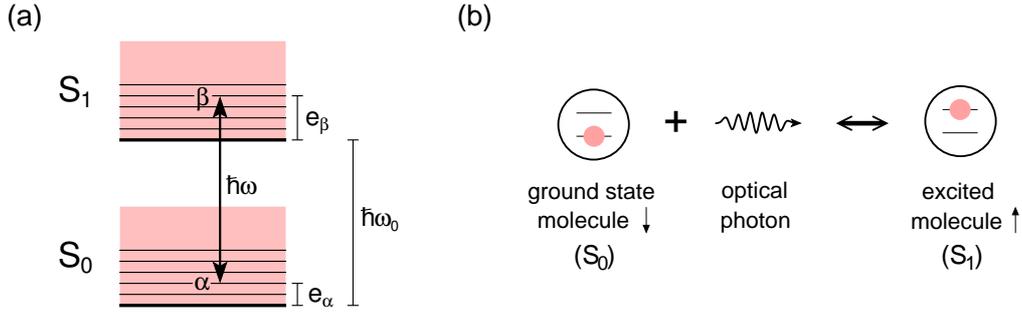}
\par\end{centering}

\begin{centering}
\vspace{3mm}

\par\end{centering}

\caption{\label{fig:Jablonski}(a) Jablonski diagram of a dye molecule with
two electronic levels. Both lower and upper electronic state, $S_{0}$
and $S_{1}$, are split into a rovibronic substructure. (b) The process
of absorption and fluorescence of a photon can be seen as a photochemical
reaction.}

\centering{}\vspace{3mm}

\end{figure}

Early experimental work has shown that spectra of dye molecules in
liquid solution show several universal properties \cite{Lakowicz:1999,Sharma:1999,Drexhage:DyeLasers1990,Schaefer:DyeLasers1990}.
These are for example the mirror rule, which states that the spectral
absorption profile is a mirror image of the fluorescence profile;
the Stokes rule, stating that the spectral centroid of fluorescence
occurs at a higher wavelength than that of absorption; and Kasha's
rule \cite{Kasha:1950}, which expresses that the fluorescence does
not depend on the wavelength of the exciting light. These common properties,
which to good accuracy are fulfilled in many dye species, suggest
that absorption and fluorescence in such systems follow a general
mechanism. Most of these properties can be understood as being the
consequence of a collisionally induced thermalization mechanism \cite{Lakowicz:1999,Sharma:1999}.
For a corresponding model, consider an idealized dye molecule with
an electronic ground state $S_{0}$ and an electronically excited
state $S_{1}$, each of which are subject to additional rovibronic
level splitting, as shown in Fig. \ref{fig:Jablonski}a. The vibrational
and rotational state of a dye molecule in liquid solution is permanently
altered by collisions with solvent molecules, e.g. on a femtosecond
timescale at room temperature. The frequent collisions lead to a thermalization
of the rovibrational state within a sub-picosecond timescale, which
is much faster than the electronic lifetime of the excited state,
being typically of the order of nanoseconds. When a photon is absorbed,
the dye molecule is likely to be transfered to a highly rovibronically
excited substate of the $S_{1}$ manifold, but the excessive rovibronic
energy will be quickly dissipated into the solvent bath. The fluorescence
photon will be emitted from a dye state that is in thermal equilibrium
with the solvent bath, with typically lower rovibrational quantum
number. This thermalization process, which occurs both in the electronically
excited and in the ground state level, explains why fluorescence can
be dissipative (Stokes shift) and why there typically is no correlation
between the wavelength of the absorbed and the emitted photon (Kasha's
rule). 

The thermalization of the rovibronic degrees of freedom has another,
though closely related, consequence: the Einstein coefficients of
absorption and emission at a certain photon energy $\hbar\omega$
are connected by the Boltzmann factor of that energy. This relation
is known as the Kennard-Stepanov law, and can be written in the form\begin{equation}
\frac{B_{21}(\omega)}{B_{12}(\omega)}=\frac{w_{\downarrow}}{w_{\uparrow}}\, e^{-\frac{\hbar(\omega-\omega_{0})}{k_{\textrm{B}}T}}\label{eq:KS}\end{equation}
where $B_{12}(\omega)$ and $B_{21}(\omega)$ are the Einstein coefficients
of absorption and stimulated emission respectively, $\omega_{0}$
is the frequency of the zero-phonon line of the dye, and $w_{\downarrow,\uparrow}$
are statistical weights related to the rovibronic density of states,
which will be defined subsequently in this text. We note that the
Kennard-Stepanov relation can also be stated in terms of $A(\omega)/B_{12}(\omega)$,
where $A(\omega)$ is the Einstein coefficient for spontaneous emission
\nobreakdash- with the difference to the above definition essentially
being the density of states. This relation has been discovered in
the beginning of the last century \cite{Kennard:1918p1291,Kennard:1927p1292},
and has been 'rediscovered' several times. A short historical outline
can be found in \cite{Ross:1967p1482}. Both theoretical and experimental
investigations can be found in the literature \cite{Stepanov:1957,Stepanov:1957_2,McCumber:1964p1463,Ross:1967p1482,VanMetter:1976p2141}. 

In the following, we give a short derivation of the Kennard-Stepanov
relation. As described above, we model the dye molecule by an electronic
two-level system with levels $S_{0}$ and $S_{1}$, also denoted by
$\downarrow$ and $\uparrow$, each of which is subject to additional
rovibronic level splitting. It is important to note that the Einstein
coefficients of such a medium at a given frequency $\omega$ are an
average over all pairs of individual rovibronic substates $(\alpha,\beta)$,
with $\alpha\in S_{0}$, $\beta\in S_{1}$ that match the transition
frequency:

\begin{equation}
e_{\alpha}+\hbar\omega=\hbar\omega_{0}+e_{\beta}\enskip\textrm{.}\label{eq:en_match}\end{equation}
The latter equation expresses energy conservation, see also the Jablonski
diagram of Fig. \ref{fig:Jablonski}a. We assume that the population
of rovibronic states in both lower and upper electronic states is
fully thermalized from frequent collisions with solvent molecules.
This assumption will be valid as long as the radiative lifetime of
the electronically excited state $S_{1}$ remains clearly longer than
the thermalization time. We thus expect that the corresponding substates
within the lower and upper electronic manifolds will be occupied with
a propability given by the Boltzmann factors\begin{equation}
p_{\alpha}=e^{-\frac{e_{\alpha}}{k_{\text{B}}T}}/w_{\downarrow}\quad\quad\text{and}\quad\quad p_{\beta}=e^{-\frac{e_{\beta}}{k_{\text{B}}T}}/w_{\uparrow}\enskip\text{,}\label{eq:pa_pb}\end{equation}
with the normalization factors\begin{equation}
w_{\downarrow}=\sum_{\alpha\in S_{0}}e^{-\frac{e_{\alpha}}{k_{\text{B}}T}}\quad\quad\text{and}\quad\quad w_{\uparrow}=\sum_{\beta\in S_{1}}e^{-\frac{e_{\beta}}{k_{\text{B}}T}}\enskip\textrm{.}\label{eq:w_discrete}\end{equation}
From eqn \eqref{eq:en_match} and \eqref{eq:pa_pb} one immediately
obtains\begin{equation}
p_{\beta}=\frac{w_{\downarrow}}{w_{\uparrow}}\, e^{-\frac{\hbar(\omega-\omega_{0})}{k_{\text{B}}T}}\, p_{\alpha}\enskip\textrm{.}\label{eq:p_beta}\end{equation}
We can now write\begin{eqnarray}
\frac{B_{21}(\omega)}{B_{12}(\omega)} & = & \frac{\sum_{(\alpha,\beta)}p_{\beta}\, B(\beta\rightarrow\alpha)}{\sum_{(\alpha,\beta)}p_{\alpha}\, B(\alpha\rightarrow\beta)}\,\text{,}\end{eqnarray}
where we have introduced the Einstein coefficients $B(\alpha\rightarrow\beta)$
for transitions between the corresponding individual rovibronic states.
Applying $B(\alpha\rightarrow\beta)=B(\beta\rightarrow\alpha)$ and
using in addition \eqref{eq:p_beta}, we finally obtain\begin{eqnarray}
\frac{B_{21}(\omega)}{B_{12}(\omega)} & = & \frac{w_{\downarrow}}{w_{\uparrow}}\, e^{-\frac{\hbar(\omega-\omega_{0})}{k_{\text{B}}T}}\,\frac{\sum_{(\alpha,\beta)}p_{\alpha}\, B(\alpha\rightarrow\beta)}{\sum_{(\alpha,\beta)}p_{\alpha}\, B(\alpha\rightarrow\beta)}\nonumber \\
 & = & \frac{w_{\downarrow}}{w_{\uparrow}}\, e^{-\frac{\hbar(\omega-\omega_{0})}{k_{\text{B}}T}}\enskip\textrm{,}\end{eqnarray}
which is the Kennard-Stepanov law of eqn \eqref{eq:KS}. As described
above, the Kennard-Stepanov relation is experimentally well established.
However, many dye species show smaller or even larger deviations.
For a discussion of these issues, see section \ref{sub:Spectral-temperature}.

\subsection{Chemical equilibrium between photons and molecules}

Molecules in the electronic ground state can be transferred to the
upper electronic level by the absorption of a photon, while photon
emission leads to de-excitation. These processes can be seen as a
photochemical reaction of the type\begin{equation}
\gamma+\downarrow\:\leftrightharpoons\:\uparrow\;\text{,}\end{equation}
see also the illustration of Fig. \ref{fig:Jablonski}b. Here $\gamma$
stands for a photon, $\downarrow$ for a molecule in the electronic
ground state, and $\uparrow$ for a molecule in the electronically
excited state. In chemical equilibrium the corresponding chemical
potentials, with $\mu_{\downarrow}$ ($\mu_{\uparrow}$) denoting
the chemical potential of a ground (excited) state molecule and $\mu_{\gamma}$
indicating the photon chemical potential, satisfy the relation \begin{equation}
\mu_{\gamma}+\mu_{\downarrow}=\mu_{\uparrow}\;.\end{equation}
If we express this relation in terms of the photon fugacity $z$,
the equation reads \begin{equation}
z=e^{\frac{\mu_{\gamma}}{k_{B}T}}=e^{\frac{\mu_{\uparrow}}{k_{B}T}}/e^{\frac{\mu_{\downarrow}}{k_{B}T}}\;\textrm{.}\label{eq:z}\end{equation}
Next, let the partition function of a dye molecule be \begin{equation}
\mathcal{F}=w_{\downarrow}\, e^{\frac{\mu_{\downarrow}}{k_{\textrm{B}}T}}+w_{\uparrow}\, e^{-\frac{\hbar\omega_{0}-\mu_{\uparrow}}{k_{\textrm{B}}T}}\;\text{,}\end{equation}
where $w_{\downarrow,\uparrow}$ are the statistical weights of eqn
\eqref{eq:w_discrete}. Further, $\omega_{0}$ is the frequency of
the zero-phonon line of the dye. We can then identify\begin{eqnarray}
w_{\uparrow}\, e^{-\frac{\hbar\omega_{0}-\mu_{\uparrow}}{k_{\textrm{B}}T}}/\mathcal{F} & = & \frac{\rho_{\uparrow}}{\rho}\;\text{,}\\
w_{\downarrow}\, e^{\frac{\mu_{\downarrow}}{k_{\textrm{B}}T}}\;\enskip\quad/\mathcal{F} & = & \frac{\rho_{\downarrow}}{\rho}\;\text{,}\end{eqnarray}
as the probability of finding a molecule in the excited (ground) state,
where $\rho_{\uparrow}$ ($\rho_{\downarrow}$) is the spatial density
of the excited (ground) state dye molecules respectively, and $\rho$
is the total dye molecular density. Using eqn \eqref{eq:z}, we find
that the photon chemical potential is determined by the relation

\begin{equation}
z=e^{\frac{\mu_{\gamma}}{k_{\text{B}T}}}=\frac{w_{\downarrow}}{w_{\uparrow}}\,\frac{\rho_{\uparrow}}{\rho_{\downarrow}}\, e^{\frac{\hbar\omega_{0}}{k_{B}T}}\quad\textrm{.}\label{eq:z2}\end{equation}
In chemical equilibrium the photon chemical potential is thus determined
by the excitation ratio $\rho_{\uparrow}/\rho_{\downarrow}$, which
corresponds to the relative number of dye molecules in the electronically
excited state \cite{Klaers:2011,Klaers2:2011}.

\subsection{\label{sub:Markov}Thermal equilibrium and Markov-processes }

To begin the discussion of the thermalization process, let us characterize
thermal equilibrium in terms of a Markov process. In general, the
time evolution of a physical system coupled to a heat bath can be
regarded as a random walk in configuration space \cite{Metropolis:1953p1516,Landau:Guide2000}.
Such a random walk is fully characterized by the present state of
the system and fixed transition rates between configurations (process
'without memory'). Not all sets of transition rates are physically
meaningful. A necessary requirement is that for a sufficiently long
time evolution of the system each configuration has to occur with
the statistical weight of its Boltzmann factor. In the following,
we derive a condition from which one can easily decide if a given
set of transition rates leads to thermal equilibrium. In the next
section this will then be applied to the fluorescence induced thermalization
process of a photon gas. 

Let $K$ denote a certain state of a physical system (for a photon
gas this will be given by a set of mode occupation numbers). Furthermore,
$p_{K}(t)$ denotes the probability to find state $K$ at time $t$.
The temporal evolution of $p_{K}(t)$ is given by the master equation
\begin{equation}
p_{K}(t+1)-p_{K}(t)=\sum_{K'}p_{K'}(t)\, R(K'\rightarrow K)-\sum_{K'}p_{K}(t)\, R(K\rightarrow K')\,\text{,}\end{equation}
where the coefficients $R(K\rightarrow K')$ denote the transition
rates between configurations. We are interested in transition rates
that asymptotically bring the system into thermal equilibrium, i.e.
which yield $p_{K}(t\rightarrow\infty)=\exp(-E_{K}/k_{\textrm{B}}T)/Z$,
with $Z$ as the partition function. The asymptotic master equation
for this case is given by \begin{equation}
0=\sum_{K'}\exp(-E_{K'}/k_{\textrm{B}}T)\, R(K'\rightarrow K)-\sum_{K'}\exp(-E_{K}/k_{\textrm{B}}T)\, R(K\rightarrow K')\,\text{.}\label{eq:asymptotic_mastereq}\end{equation}
Equation \eqref{eq:asymptotic_mastereq} has many solutions. One solution
stands out by showing no net probability flow between two given states
(detailed balance): \begin{equation}
0=\exp(-E_{K'}/k_{\textrm{B}}T)\, R(K'\rightarrow K)-\exp(-E_{K}/k_{\textrm{B}}T)\, R(K\rightarrow K')\:\:\forall\, K,K'\,\,\text{,}\end{equation}
or\begin{equation}
\frac{R(K\rightarrow K')}{R(K'\rightarrow K)}=\exp(-\Delta E/k_{\textrm{B}}T)\:\:\:\:\forall\, K,K'\:\:\text{,}\label{eq:detailed_balance}\end{equation}
with an energy difference $\Delta E=E_{K'}-E_{K}$. If the transition
rates $R(K\rightarrow K')$ fulfill eqn \eqref{eq:detailed_balance},
this gives a sufficient condition that the Markov process drives the
system into thermal equilibrium.

\subsection{Light-matter thermalization process}

\begin{figure}
\begin{centering}
\vspace{0mm}

\par\end{centering}

\begin{centering}
\includegraphics[width=0.25\columnwidth]{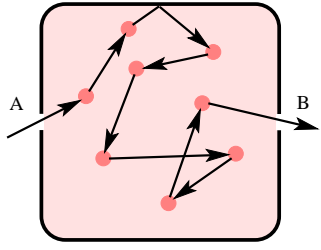}
\par\end{centering}

\begin{centering}
\vspace{3mm}

\par\end{centering}

\caption{\label{fig:dye_box}Multiple absorption-fluorescence cycles in a macroscopic
dye-filled photon box. Asymptotically, the photons entering the box at point A would be transferred to blackbody radiation at the temperature of the environment. However, without a low-frequency cutoff the thermalization process remains incomplete, as photons leave the absorption bandwidth of the dye.}

\centering{}\vspace{3mm}

\end{figure}

To begin with, consider an experimental scenario as depicted in Fig.
\ref{fig:dye_box}, consisting of a macroscopic box with reflecting
walls that is filled with a dye solution.%
Light, emitted e.g. by a spectrally narrow optical source, is irradiated
at point A. What is the spectrum of the radiation leaving the dye
container at point B? If the optical density of the dye is sufficiently
large, multiple absorption-emission fluorescence cycles will occur.
One then experimentally finds that the spectrum at point B is red
shifted with respect to that obtained for a single fluorescence event.
In the context of fluorescence spectroscopy, this effect is known
as the inner filter effect \cite{Lakowicz:1999,Sharma:1999}. Interestingly,
this red shift of the fluorescence can be understood in terms of a
partial thermalization process. By fluorescence and reabsorption,
the photons exchange energy with the dye solution. This energy exchange
establishes a thermal contact between the light and the dye medium.
In principle, one would expect that the state of the light field asymptotically
relaxes towards a thermal state at the temperature of the dye solution,
i.e. after many complete cycles the spectrum of blackbody radiation
at room temperature would be obtained. However, owing to the succesive
red shift, along with the limited spectral bandwith of the dye, the
reabsorption probability rapidly decreases, and the dye solution gradually
becomes transparent. The thermal contact soon breaks down and the
process remains incomplete. Such an incomplete thermalization occurs
e.g. in some solar light concentrators \cite{Meyer:2009p2105}. 

Regarding photon number conservation, it on the other hand is not
desirable to transfer the irradiated light into blackbody radiation.
As soon as the photon energy becomes comparable to thermal energy,
one would lose the ability to experimentally tune the photon number.
In equilibrium, the photon number would then readjust following the
available thermal energy, independently of the optical pumping power
irradiated at point A (provided that the pumping does not significantly
heat up the dye solution). However, in a microcavity environment,
as it is used in our experiment, both a full thermalization and an
experimentally tunable photon number can be achieved. This becomes
clear by noting that the low-frequency cutoff, imposed by the resonator,
prevents a succesive red-shift of the photon gas. The thermalization
by repeated absorption-emission cycles can now proceed until the photon
gas is fully thermalized, i.e. thermal contact to the dye solution
is not lost. Owing to the low-frequency cutoff, the photon energy
can remain far above the thermal energy. With $\hbar\omega_{\text{cutoff}}\simeq2.1\text{eV}\gg1/40\,\text{eV}\simeq k_{\text{B}}T$
this limit is fulfilled in our experiment. In this situation, no photons
are created or destroyed on average. The thermalization process here
conserves the photon number, with the latter being determined by the
strength of the optical pumping. 

In the remainder of this section, we present a more rigorous treatment
of the light-matter thermalization \cite{Klaers:2010p2137,Klaers2:2011}.
The thermalization process is considered as a random walk in the configuration
space of all allowed light field states. Here a state $K$ is given
by the cavity mode occupation numbers $K=(n_{0}^{K},n_{1}^{K},n_{2}^{K},\ldots)$.
The mode occupation numbers are permanently altered by photon absorption
and emission processes. In first order perturbation theory, the rates
(per volume) for absorption and emission of one photon in mode $i$
at cavity position $\mathbf{r}$, denoted by $R_{12}^{K,i}(\mathbf{r})$
and $R_{21}^{K,i}(\mathbf{r})$, have the form \begin{eqnarray}
R_{12}^{K,i}(\mathbf{r}) & = & B_{12}(\omega_{i})\, u^{i}(\mathbf{r})\,\rho_{\downarrow}\, n_{i}^{K}\label{eq:R12}\\
R_{21}^{K,i}(\mathbf{r}) & = & B_{21}(\omega_{i})\, u^{i}(\mathbf{r})\,\rho_{\uparrow}\,(n_{i}^{K}+1)\label{eq:R21}\end{eqnarray}
where $u^{i}(\mathbf{r})$ is the spectral energy density of one photon
in mode $i$. We assume that the number of dye molecules is sufficiently
large, such that photon absorption and emission leave the densities
of ground state ($\rho_{\downarrow}$) and excited state ($\rho_{\uparrow}$)
molecules unchanged, i.e. $\rho_{\uparrow,\downarrow}$ can be treated
as fixed parameters. This assumption corresponds a grandcanonical
limit, see also section \ref{sub:Condensate-fluctuations}.

Suppose that a state $K'$ emerges from state $K$ by the absorption
of a photon in mode $i$, with $n_{i}^{K'}=n_{i}^{K}-1$, and accordingly,
that $K$ emerges from $K'$ by an emission process into this mode.
The corresponding rates are $R_{12}^{K,i}(\mathbf{r})$ and $R_{21}^{K',i}(\mathbf{r})$,
and with eqn \eqref{eq:R12} and \eqref{eq:R21}, their ratio is given
by $R_{12}^{K,i}(\mathbf{r})/R_{21}^{K',i}(\mathbf{r})=B_{12}(\omega_{i})\,\rho_{\downarrow}/B_{21}(\omega_{i})\,\rho_{\uparrow}$.
In section \ref{sub:Markov} it was shown that the random walk given
by the transition rates of eqn \eqref{eq:R12} and \eqref{eq:R21}
will lead to thermal equilibrium, if the ratio $R_{12}^{K,i}(\mathbf{r})/R_{21}^{K',i}(\mathbf{r})$
is given by the Boltzmann factor of the energy difference between
$K$ and $K'$ \begin{equation}
\frac{B_{12}(\omega_{i})}{B_{21}(\omega_{i})}\,\frac{\rho_{\downarrow}}{\rho_{\uparrow}}\overset{!}{=}e^{\frac{\hbar\omega_{i}-\mu_{\gamma}}{k_{\textrm{B}}T}}\enskip\textrm{,}\label{eq:DB}\end{equation}
see also \cite{Norman:1969p208,Klaers2:2011,Klaers:Thesis2011}. If
we now apply the Kennard-Stepanov relation, eqn \eqref{eq:KS}, and
assume chemical equilibrium, eqn \eqref{eq:z2}, the detailed balance
condition of eqn \eqref{eq:DB} is indeed verified. This means that
multiple absorption-emission cycles drive the photon gas into thermal
equilibrium with the dye solution at temperature $T$, and with a
photon chemical potential $\mu_{\gamma}$ determined by the molecular
excitation ratio. The average occupation number $\bar{n}_{i}$ of
mode $i$ can be determined by balancing the average absorption and
emission rates at a given cavity position. As expected, this gives
a Bose-Einstein distribution for the average occupation number, with
$\bar{n}_{i}\hspace{-0.7mm}=\hspace{-0.7mm}\bigl(\exp\left[\left(\hbar\omega_{i}\hspace{-0.7mm}-\hspace{-0.7mm}\mu_{\gamma}\right)\hspace{-0.7mm}/k_{\textrm{B}}T\right]\hspace{-0.7mm}-\hspace{-0.7mm}1\bigr)^{-1}$.

\subsection{\label{sub:Spectral-temperature}Spectral temperature }

From the Kennard-Stepanov theory of dye spectra, as presented in section
\ref{sub:Kennard-Stepanov}, one expects that the spectra of absorption
and emission are interlinked by a Boltzmann factor. In general, this
law is empirically well confirmed. However, it is known that many
real dyes show more or less pronounced deviations. Before choosing
a particular dye medium, it is thus necessary to check whether it
fulfills the Kennard-Stepanov relation. For this it proves helpful
to introduce a spectral temperature, which is derived using a formal
solution of eqn \eqref{eq:KS}. Consider the Kennard-Stepanov relation
for two different frequencies $\omega$ and $\omega+\delta\omega$.
From eqn \eqref{eq:KS}, we obtain\begin{equation}
\frac{B_{21}(\omega)}{B_{12}(\omega)}\frac{B_{12}(\omega+\delta\omega)}{B_{21}(\omega+\delta\omega)}=e^{\frac{\hbar\delta\omega}{k_{\text{B}}T}}\,\text{.}\end{equation}
We can now solve this equation for the frequency dependent spectral
temperature $T_{\text{spec}}(\omega)$:\begin{eqnarray}
T_{\text{spec}}(\omega) & = & \frac{\hbar\delta\omega}{k_{\text{B}}\,\ln\left(\frac{B_{21}(\omega)}{B_{12}(\omega)}\frac{B_{12}(\omega+\delta\omega)}{B_{21}(\omega+\delta\omega)}\right)}\label{eq:T_spec}\\
 & \overset{\delta\omega\rightarrow0}{=} & \frac{\hbar}{k_{\text{B}}\,\frac{\partial\ln}{\partial\omega}\frac{B_{12}(\omega)}{B_{21}(\omega)}}\,\text{,}\label{eq:T_spec2}\end{eqnarray}
and eventually perform the limit $\delta\omega\rightarrow0$. Thus,
if the spectral profiles of absorption and emission are known, one
can easily calculate the corresponding spectral temperature. We note
that the particular form of $T_{\text{spec}}$ given by eqn \eqref{eq:T_spec}
and \eqref{eq:T_spec2} has the advantage of being independent of
the statistical weights $w_{\downarrow,\uparrow}$ and the frequency
of the zero-phonon line $\omega_{0}$, which might not be known in
all cases. Moreover, in this form it is sufficient to know the coefficients
$B_{12,21}(\omega)$ only up to a proportionality factor.

\begin{figure}
\noindent \begin{centering}
\vspace{5mm}

\par\end{centering}

\noindent \begin{centering}
\includegraphics[angle=270,scale=0.55]{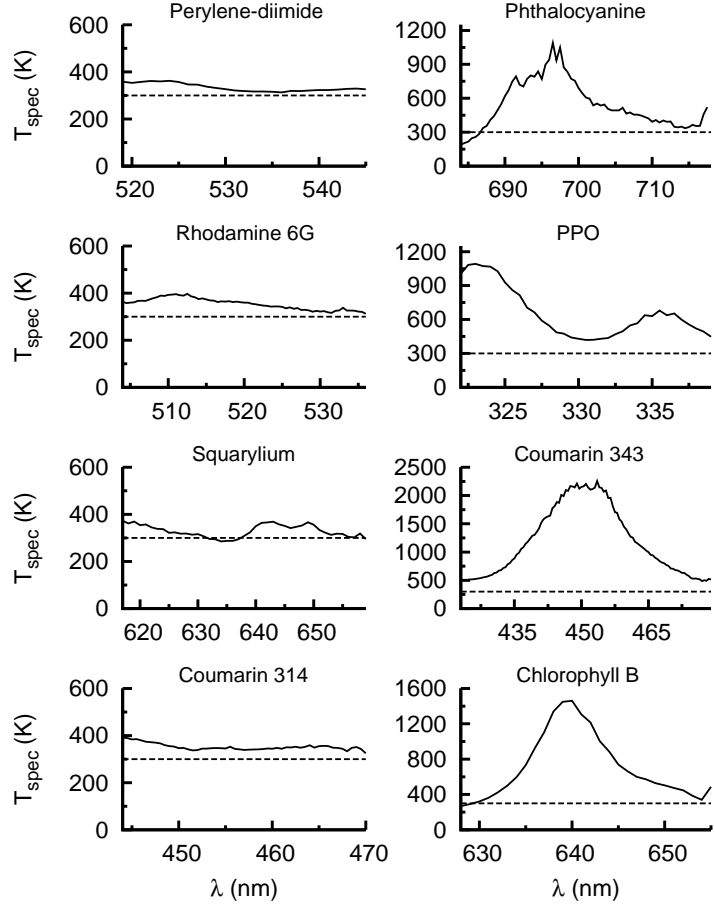}\vspace{5mm}

\par\end{centering}

\caption{\selectlanguage{ngerman}%
\label{fig:Spektrale Temperaturen}\foreignlanguage{english}{Spectral
temperatures $T_{\textrm{spec}}(\lambda)$ for different dyes at room
temperature, derived from the spectra of the database of Ref. \cite{Du:1998}.
This numerical determination is most precise in the Stokes region,
where both absorption and fluorescence are sufficiently strong and
thus most precisely known. We nevertheless estimate an error of the
given spectral temperatures of at least $15\,\text{\%}$. The corresponding
quantum efficiencies are: perylene-diimide: $\Phi=0.97$, rhodamine
6G: $\Phi=0.95$, squarylium: $\Phi=0.65$, coumarin 314: $\Phi=0.68$,
phthalocyanine: $\Phi=0.60$, PPO: $\Phi\approx1$, coumarin 343:
$\Phi=0.63$, chlorophyll B: $\Phi=0.117$. }\selectlanguage{english}
}

\noindent \centering{}\vspace{5mm}

\end{figure}

The Kennard-Stepanov relation holds for a given dye, if its spectral
temperature coincides with the thermodynamic temperature of the solution,
$T_{\text{spec}}(\omega)=T$, independent of the frequency $\omega$.
Figure \ref{fig:Spektrale Temperaturen} shows the spectral temperatures,
here as function of the wavelength, for eight different dye species.
They were calculated with eqn. \eqref{eq:T_spec} using the corresponding
spectra of the database of Ref. \cite{Du:1998}. The dyes on the left
hand side of the figure fulfill the condition $T_{\text{spec}}(\omega)\simeq T$
to good approximation, while the dyes on the right hand side show
significant deviations. Noteworthy, the spectral temperature here
has a tendency to be higher than the ambient temperature, see also
\cite{Kozma:ActaPhys1964,VanMetter:1976p2141}. As explanation for
the observed deviations, one mainly finds two lines of reasoning in
the literature, either based on incomplete rovibronic thermalization
or on inhomogeneous broadening \cite{Mazurenko:OptSpectr1972,Mazurenko:OptSpectr1974,VanMetter:1976p2141,Sawicki:1996p2109}.
Our experiment uses either perylene-dimide or rhodamine 6G dye, for
both of which the quantum efficiency is $95\%$ or above, and the
Kennard-Stepanov relation is well fulfilled. %

\section{Thermodynamics of the two-dimensional photon gas }

Formally, the photon gas confined in the resonator is equivalent to
a two-dimensional gas of massive bosonic particles. This can be seen
by investigating the energy momentum relation. We will derive the
equilibrium properties of such a gas, and discuss the expected condensate
fluctuations.

\subsection{Cavity photon dispersion }

We begin by expressing the energy of a photon in the microcavity as
a function of longitudinal (i.e. along the optical axis) wavenumber
$k_{z}$ and transverse wavenumber $k_{r}$, \begin{equation}
E=\frac{\hbar c}{n_{0}}|k|=\frac{\hbar c}{n_{0}}\sqrt{k_{z}^{2}+k_{r}^{2}}\,\text{,}\end{equation}
where $n_{0}$ is the refractive index of the medium. Let us denote
the spacing between the two cavity mirrors on the optical axis as
$D_{0}$ ($\simeq1.46\,\text{\textmu m}$ in our experiment), and
the radii of the two curved mirrors, which we assume to be identical,
as R ($\simeq1\,\text{m}$ typically). The boundary conditions generated
by the mirrors imply a resonance condition \begin{equation}
k_{z}(r)=q\pi/D(r)\,\text{,}\end{equation}
with $q$ as the longitudinal mode number and $D(r)=D_{0}-2(R-\sqrt{R^{2}-r^{2}})$
as the mirror spacing at a distance $r$ from the optical axis. In
a paraxial limit, with $k_{r}\ll k_{z}$ and $r\ll R$, we can approximate
the photon energy by the expansion \begin{equation}
E\simeq\frac{m_{\text{eff}}c^{2}}{n_{0}^{2}}+\frac{(\hbar k_{r})^{2}}{2m_{\text{eff}}}+\frac{1}{2}m_{\text{eff}}\Omega^{2}r^{2}\,\text{,}\end{equation}
with $m_{\text{eff}}=\hbar k_{z}(0)n_{0}/c=\hbar\omega_{\text{cutoff}}n_{0}^{2}/c^{2}$
and $\Omega=(c/n_{0})/\sqrt{D_{0}R/2}$. This yields the dispersion
relation of a particle with mass $m_{\text{eff}}$ moving in the transverse
resonator that is subject to additional harmonic confinement with
trapping frequency $\Omega$. If we in addition account for a nonlinear
self-interaction of photons that modifies the refractive index following
$n_{0}\rightarrow n_{0}+n_{2}I(r)$ , where $n_{2}$ is a nonlinear
index of refraction, and $I(r)$ is the optical intensity, we find
that the photon energy is shifted by an amount \begin{equation}
E_{\text{int}}\simeq-m_{\text{eff}}c^{2}\frac{n_{2}}{n_{0}^{3}}I(r)\,\text{.}\label{eq:E_int}\end{equation}
A non-zero value of $n_{2}$ can arise e.g. from a Kerr nonlinearity
or thermal lensing. Since the confined photons have the quadratic
dispersion relation of massive particles, we can define a thermal
de Broglie wavelength by $\lambda_{\text{th}}=h/\sqrt{2\pi m_{\text{eff}}k_{\text{B}}T}$,
in direct analogy to e.g. a gas of atoms. Essentially, $\lambda_{\text{th}}$
is inversely proportional to the average transversal wavenumber $\sqrt{\left\langle k_{r}^{2}\right\rangle }$
of the photons, and its value at room temperature is $\simeq1.58\,\text{\textmu m}$.
Bose-Einstein condensation is expected when the phase space density
$n\lambda_{\text{th}}^{2}$ exceeds a value near unity, where $n$
denotes the two-dimensional number density.

\subsection{Statistical theory of two-dimensional Bose gas in trap}

For an exact determination of the phase transition, a statistical
multimode treatment is necessary. The average photon number at a transversal
energy $u=\hbar(\omega-\omega_{\text{cutoff}})$ is given by a Bose-Einstein
distribution \begin{equation}
n_{T,\mu}(u)=\frac{g(u)}{e^{\frac{u-\mu}{k_{\text{B}}T}}-1}\,\text{,}\label{eq:BEdistr}\end{equation}
where $g(u)=2(u/\hbar\Omega+1)$ is the degeneracy factor (which here
increases linearly with energy), and the factor $2$ accounts for
the two possible polarizations. A macroscopic occupation of the ground
state mode at $u=0$ sets in when the particle number $N$ reaches
a critical value of $N_{c}=\sum_{u>0}n_{T,\mu=0}(u)$. We find \begin{equation}
N_{c}=\frac{\pi^{2}}{3}\left(\frac{k_{\text{B}}T}{\hbar\Omega}\right)^{2}\,\text{.}\label{eq:Nc}\end{equation}
The typical photon trapping frequency in our setup is $\Omega\simeq2\pi\cdot4.1\times10^{10}\,\text{Hz}$.
For room temperature, $T=300\,\text{K}$, we arrive at $N_{\text{c}}\simeq77000$,
which is experimentally feasible. If this critical value is reached,
the occupation of transversally excited modes is expected to saturate
and the ground mode starts to become macroscopically populated. The
possibility to observe a Bose-Einstein condensation at room temperature
can be understood from the extremely small effective photon mass $m_{\text{eff}}=\hbar\omega_{\text{cutoff}}n_{0}^{2}/c^{2}$,
which in our case corresponds to $2.1\,\text{eV}\, n_{0}^{2}/c^{2}\simeq7\cdot10^{-36}\,\text{kg}$.
This is $10$ orders of magnitude below the mass of alkali atoms,
which extremely enhances the phase transition temperature with respect
to atomic systems.

\subsection{\label{sub:Condensate-fluctuations}Condensate fluctuations }

In our experiment, photons are initially brought into the system by
pumping the dye medium with a laser. The pumping is maintained throughout
the measurement to compensate for photon losses due to unconfined
optical modes, finite dye quantum efficiency and mirror losses. Despite
pumping, the photon gas is expected to both spectrally and spatially
relax to thermal equilibrium, provided that a photon scatteres several
time off a dye molecule before being lost \cite{Klaers:2010p2137}.
By optical pumping we generate and uphold a reservoir of electronic
excitations in the dye medium that can exchange particles with the
photon gas. %
\begin{figure}
\begin{centering}
\vspace{0mm}

\par\end{centering}

\begin{centering}
\includegraphics[width=0.3\columnwidth]{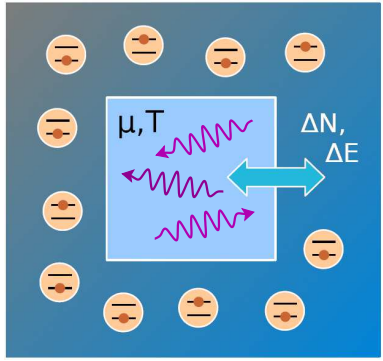}
\par\end{centering}

\begin{centering}
\vspace{3mm}
\caption{\label{fig:grandcanonical}Schematic illustration of the excitation
exchange between the photon gas and a reservoir of electronically
excited dye molecules. The photon gas can be seen as an open system,
in the sense of grandcanonical experimental conditions.}

\par\end{centering}

\centering{}\vspace{5mm}

\end{figure}
This experimental situation is well described by a grandcanonical
statistical ensemble, see Fig. \ref{fig:grandcanonical}, in which
the photon gas acquires the temperature of the dye solution and the
photon chemical potential is directly related to the excitation level
in the medium, as given by eqn \eqref{eq:z2}. 

The particle exchange with the reservoir of excited state dye molecules
is of special relevance for the expected photon statistics and second
order coherence of the condensate. It is well known that there is
no ensemble equivalence for Bose-Einstein condensation \cite{Fujiwara:1970p993,ZIFF:1977p513,Kocharovsky:2006p985}.
In particular, the particle number fluctuations of the condensate
strongly depend on the underlying statistical ensemble. For a (micro-)canonical
ensemble, which is typically realized in atomic BEC experiments, the
particle number distribution of the ground state changes from Bose-Einstein-
to Poisson-like when the condensation sets in. This is accompanied
by a damping of the number fluctuations. On the other hand, the particle
number distribution for a grandcanonical ensemble remains Bose-Einstein-like
and the number fluctuations remain of the order of the average occupation
number, $\Delta n_{0}=\sqrt{\left\langle n_{0}^{2}\right\rangle -\bar{n}_{0}^{2}}\simeq\bar{n}_{0}$.
This can also be seen in the second order coherence of the condensate.
The zero-delay autocorrelation function $g^{(2)}(0)=\left\langle n_{0}(n_{0}-1)\right\rangle /\left\langle n_{0}\right\rangle ^{2}$
is not expected to drop off to $g^{(2)}(0)=1$, as it does for (micro-)canonical
BECs. Instead, a bunching behavior with $g^{(2)}(0)=2$ is expected
even below the critical temperature \cite{Klaers2:2011}. In the future,
it will be very interesting to test for such unusually large condensate
fluctuations. This regime is not observed in Bose-Einstein condensates
of ultracold atomic gases, but likely to occur for photonic BECs.

\section{Experiments on photon gas thermalization }

\subsection{Experimental setup }

\begin{figure}
\begin{centering}
\vspace{0mm}

\par\end{centering}

\begin{centering}
\includegraphics[width=0.75\columnwidth]{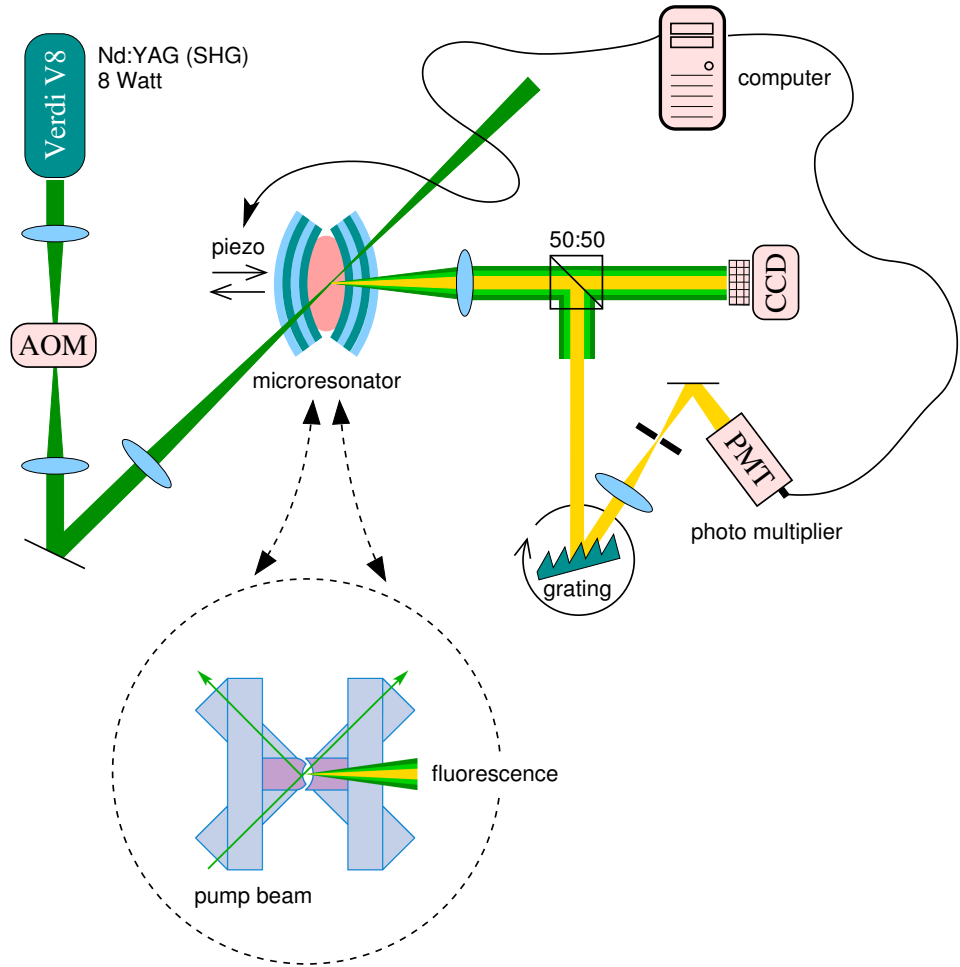}
\par\end{centering}

\begin{centering}
\vspace{3mm}
\caption{\label{fig:Setup}Schematic set-up of the microcavity experiment.
The dye-filled microresonator is pumped at an angle of 45\textdegree{}
to the optical axis. The radiation emerging from the microresonator
is examined both spatially and spectrally.}

\par\end{centering}

\centering{}\vspace{3mm}

\end{figure}

\begin{figure}
\begin{centering}
\vspace{0mm}

\par\end{centering}

\begin{centering}
\includegraphics[width=0.85\columnwidth]{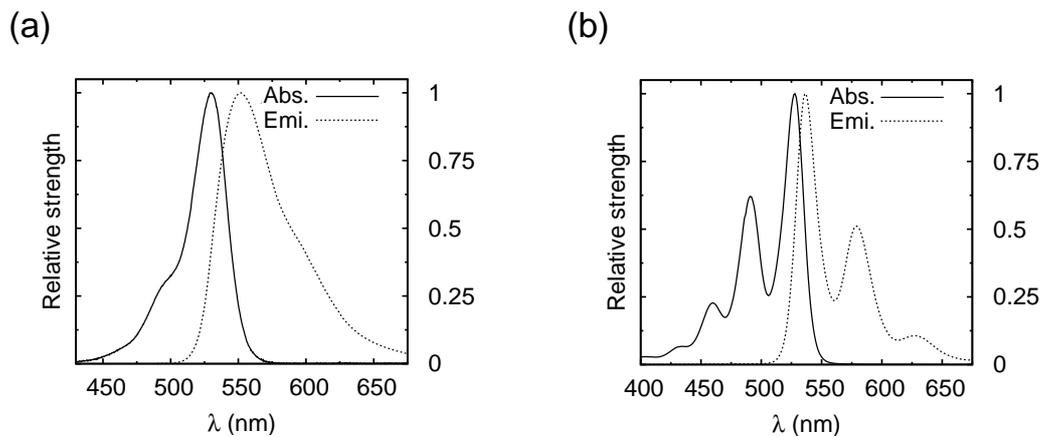}
\par\end{centering}

\begin{centering}
\vspace{0mm}
\caption{\label{fig:Rh6GPDI}Relative strengths of absorption, $B_{12}(\lambda)/B_{12}^{\text{max}}$,
and emission, $B_{21}(\lambda)/B_{21}^{\text{max}}$, for the dyes
(a) rhodamine 6G and (b) perylene diimide.}

\par\end{centering}

\centering{}\vspace{3mm}

\end{figure}

We begin by describing measurements on the thermalization of the photon
gas in the dye-filled microresonator, which have been carried out
at comparatively small photon numbers, i.e. far below the critical
photon number. A scheme of the experimental setup used in the experiments
of Refs. \cite{Klaers:2010,Klaers:2010p2137} of our group is shown
in Fig. \ref{fig:Setup}.%
 The optical resonator consists of two highly reflecting mirrors with
reflectivity above 99.997\% in the wavelength region $500-590\,\text{nm}$.
This reflectivity translates to a finesse of $>100000$ for the empty
cavity. The mirrors are spherically curved with a typical radius of
curvature of $R=1\,\text{m}$. To allow for a cavity length in the
micrometer regime despite the curvature, one of the mirrors is cut
to $1\times1\,\text{mm}$ surface size. Both resonator mirrors are
glued onto glass substrates. Additional glass prisms are attached
to the side to allow for a pumping under an angle of 45\textdegree{},
see Fig. \ref{fig:Setup}. The typical (effective) cavity length of
$D_{0}\simeq1.46\,\text{\textmu m}$ is determined from the resonator
free spectral range, and corresponds to the $q=7$ longitudinal mode.
The used dyes are either rhodamine 6G or perylenedimide (PDI), whose
absorption and fluorescence spectra are shown in Fig. \ref{fig:Rh6GPDI}
.%
The dyes have a high quantum efficiency between 95\% and 97\% \cite{Magde:2002p2101},
and spectral temperatures close to the thermodynamic temperature of
the dye solution. We use organic solvent, e.g. methanol and ethylene
glycol, and typical dye concentrations of $1.5\times10^{-3}\,\text{Mol/l}$
for rhodamine 6G. A careful filtering of the dye solution is necessary
before placing it between the mirrors. The microcavity is pumped with
a laser beam at a wavelength of $532\,\text{nm}$ derived from a frequency
doubled Nd:YAG laser. The light transmitted through one of the cavity
mirrors is split into two partial beams by a non-polarizing beamsplitter.
One beam is imaged onto a CCD-camera for spatial analysis, while the
other is sent into a spectrometer for a monitoring of the emitted
spectrum. We use different types of spectrometers, both commercially
available and self-built devices. Care has to be taken when the spectrometer
makes use of an entrance slit. The coupling efficiency to the spectrometer
will be different for the various transversally excited cavity modes.
This problem can be overcome by placing a diffusing plate in front
of the entrance slit. However, a more advantageous method is to remove
the entrance slit altogether. This is possible because the microresonator
fluorescence can be sufficiently well collimated even without an additional
spatial filtering. In Fig. \ref{fig:Setup}, a self-built version
of such a device is shown. The shown spectrometer uses a rotating
grating and a photomultiplier tube.

\subsection{Spectral and spatial intensity distribution}

\begin{figure}
\begin{centering}
\vspace{0mm}

\par\end{centering}

\begin{centering}
\includegraphics[width=1.03\columnwidth]{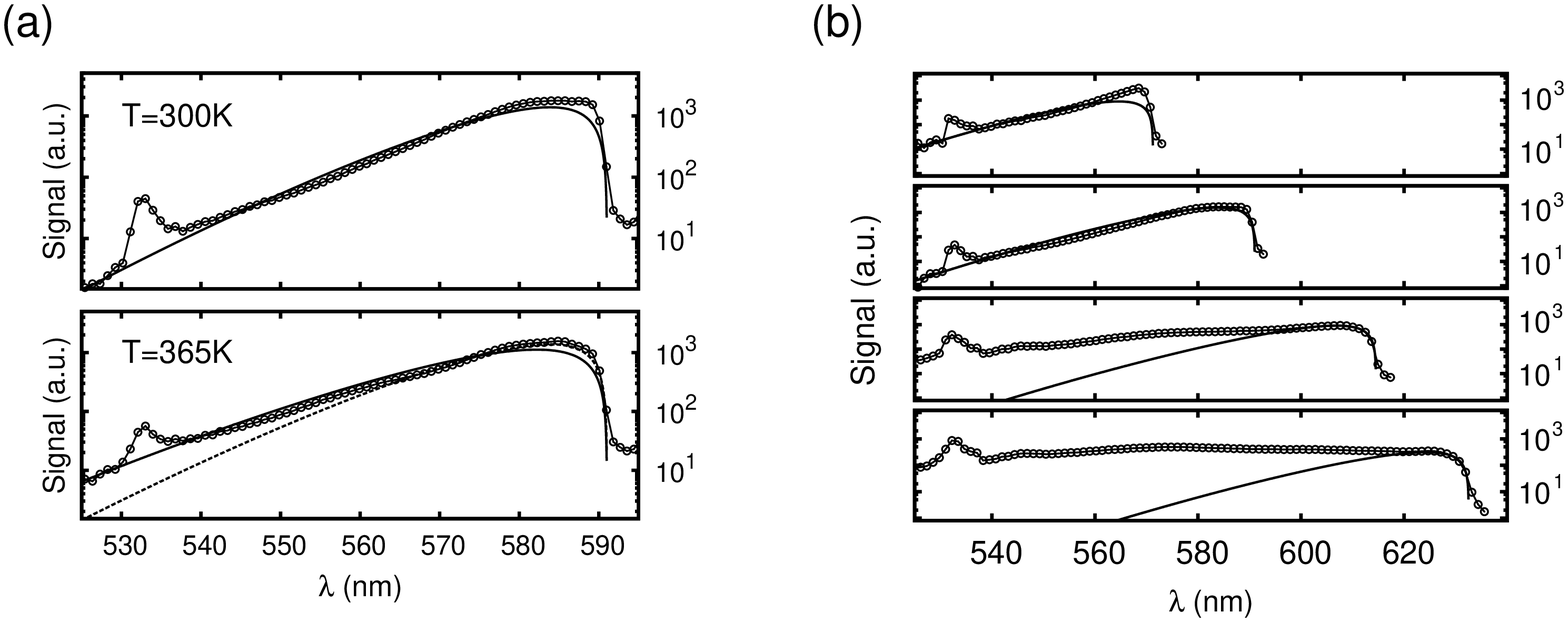}
\par\end{centering}

\begin{centering}
\vspace{0mm}
\caption{\label{fig:thermal_spectra}(a) Spectral distribution of the radiation
emitted from the microcavity far below the critical photon number
at $T=300\,\text{K}$ and $T=365\,\text{K}$ (circles). The spectra
are in good agreement with the Boltzmann-distributed photon energies
(lines). For comparison, a $T=300\,\text{K}$ Boltzmann-distribution
is additionally plotted in the bottom graph (dashed line). (Rhodamine
6G in ethylene glycol, concentration $5\times10^{-4}\,\text{mol/l}$,
mirror curvatures $R_{1}=R_{2}=1\,\text{m}$, cavity order $q=7$)
(b) Normalized spectral distribution of the radiation emitted from
the microcavity (connected circles) for 4 different cutoff wavelengths
$\lambda_{\text{c}}\simeq\{570,590,615,630\}\,\text{nm}$, from top
to bottom. In addition, Boltzmann-distributed photon energies for
$T=300\,\text{K}$ are plotted (lines).}

\par\end{centering}

\centering{}\vspace{3mm}

\end{figure}

In initial experiments, we have tested for the thermalization of the
two-dimensional photon gas in the dye-filled resonator \cite{Klaers:2010p2137}.
Figure \ref{fig:thermal_spectra}a shows spectra of light transmitted
through one of the cavity mirrors for two different temperatures of
the resonator setup (top: room temperature, $T=300\,\text{K}$, bottom:
$T=365\,\text{K}$).%
The output power for these measurements was $(50\pm5\,\text{nW})$,
which corresponds to an average photon number in the cavity of $N_{\text{ph}}=60\pm10$.
From this, the chemical potential can be determined by numerical solving
$\sum_{u\ge0}n_{T,\mu}(u)=N_{\text{ph}}$, with $n_{T,\mu}(u)$ given
by eqn \eqref{eq:BEdistr}. For the two data sets in Fig. \ref{fig:thermal_spectra}a,
we obtain $\mu/k_{\text{B}}T=-6.76\pm0.17$ ($T=300\,\text{K}$) and
$\mu/k_{\text{B}}T=-7.16\pm0.17$ ($T=365\,\text{K}$), respectively.
Clearly, both measurements are performed far below the phase transition,
which for this finite size system is expected to occur once the chemical
potential becomes comparable to the trap level spacing (i.e. $\mu\simeq-\hbar\Omega$).
Because of $-\mu/k_{\text{B}}T\gg1$, the term $-1$ in the denominator
of eqn \eqref{eq:BEdistr} can be neglected and the distribution becomes
Boltzmann-like. As can be seen from the figure, besides a derivation
near $532\,\text{nm}$ from residual pump light, the measured spectra
are in good agreement with theoretical expectations over a wide spectral
range for the two temperatures, respectively. We interpret this as
evidence for the photon gas to be in thermal equilibrium with the
dye solution. In another set of measurement, the cutoff wavelength
$\lambda_{\text{cutoff}}=2\pi c/\omega_{\text{cutoff}}$ was varied,
as can easily be achieved in our setup by a piezo tuning of the cavity
length. Corresponding spectra are shown in Fig. \ref{fig:thermal_spectra}b,
along with theory spectra obtained by assuming a Boltzmann distribution
of the transversally excited modes. Satisfactory agreement between
theory and experiment is only obtained for the two upper spectra with
cutoff wavelengths near $570\,\text{nm}$ and $590\,\text{nm}$. For
the two lower spectra with longer cutoff wavelength, no satisfactory
thermalization is present, which can be attributed to the weak dye
reabsorption in this spectral regime. These observations illustrate
the importance of both emission and reabsorption for the thermalization
process. 

We have also monitored the spatial distribution of the photon gas
at similarly low pumping power. As before, the experimental data can
be well explained by assuming a Boltzmann-like population of the transversally
excited cavity modes \cite{Klaers:2010p2137}. Furthermore, a spatial
concentration of light into the center was observed, which can be
seen as consequence of minimizing the photon energy in the effective
trapping potential induced by the mirror curvature. Technical applications
of this light concentration effect could include the collection of
diffuse solar light to a central spot \cite{VanSark:2008p727,Meyer:2009p2105}.

\section{Experiments on photon Bose-Einstein condensation}

In subsequent experiments, the dye microresonator was operated at
higher pump powers, allowing for photon numbers near and above the
critical regime. To avoid population of triplet states and excessive
heat deposition, the pump beam in these measurements was acousto-optically
chopped to $0.5\,\text{\textmu s}$ long pulses with $8\,\text{ms}$
repetition time. Since the pulses are two orders of magnitude longer
than the lifetime of the excited dye molecules and four orders of
magnitude longer than the lifetime of the photons (average time between
emission and reabsorption), the experimental conditions can be considered
as quasi-static.

\begin{figure}
\begin{centering}
\vspace{0mm}

\par\end{centering}

\begin{centering}
\includegraphics[width=0.8\columnwidth]{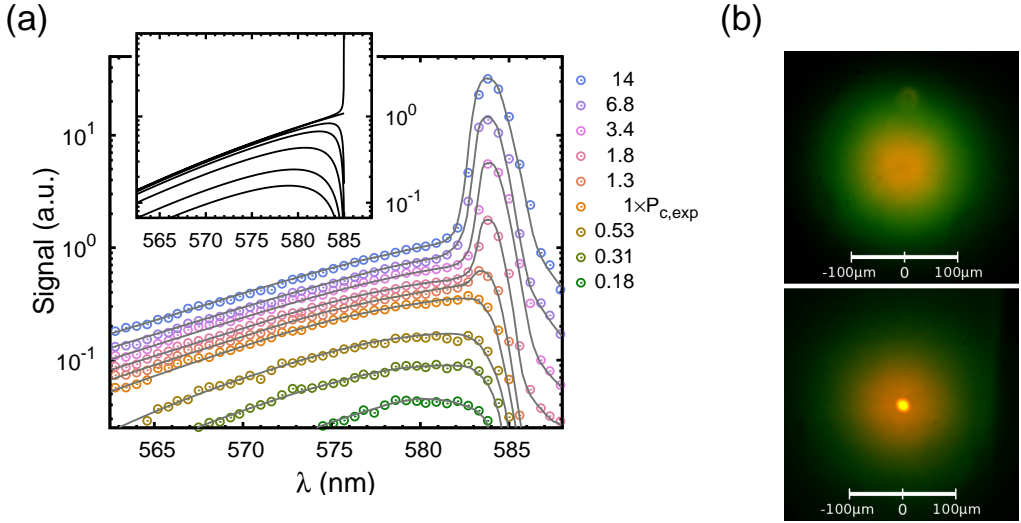}
\par\end{centering}

\begin{centering}
\vspace{3mm}
\caption{\label{fig:spectra_condensed}(a) The connected circles give measured
spectral intensity distributions for different optical pump powers.
The legend gives the intracavity optical power, determining the photon
number, in units of $P_{\text{c,exp}}=(1.55\pm0.6)\,\text{W}$. A
spectrally sharp BEC peak at the position of the cavity cutoff is
visible above the critical power. The observed peak width is limited
by the spectrometer resolution. The inset gives theoretical spectra.
(b) Images of the radiation emitted along the cavity axis, below (top)
and above (bottom) the critical power. In the latter case, a condensate
peak is visible in the center. }

\par\end{centering}

\centering{}\vspace{5mm}

\end{figure}

Typical experimental spectra of the photon gas are shown in Fig. \ref{fig:spectra_condensed}a
for different optical pumping powers \cite{Klaers:2010}.%
At low pumping and correspondingly low intracavity optical power,
a broad spectral distribution resembling a Boltzmann-like occupation
of modes above the cavity cutoff is seen. Near the phase transition,
the spectral maximum shifts towards the cutoff. When exceeding the
critical photon number, a spectrally sharp peak at the position of
the cavity cutoff is visible. The observed spectral width is limited
by the resolution of the used spectrometer. The experimental results
are in good agreement with theoretical spectra based on a Bose-Einstein-like
population of cavity modes, see the inset of the figure. The observed
imperfect saturation of the transversally excited modes for even higher
photon numbers is attributed to an interaction-induced deformation
of the effective trapping potential \cite{Tammuz:2011p230401}. At
the phase transition, the optical intracavity power is $P_{\text{c,exp}}=(1.55\pm0.6)\,\text{W}$,
which corresponds to a photon number of $(6.3\pm2.4)\times10^{4}$.
This value is in good agreement with the theoretical prediction of
eqn \eqref{eq:Nc}. Notably, a similar measurement with PDI gives
the same result within the experimental uncertainties. Thus, the obtained
results seem to be largely independent of the used dye. 

Figure \ref{fig:spectra_condensed}b shows spatial images of the radiation
emitted by the dye microcavity (real image onto a color CCD camera)
both below (top) and above (bottom) the critical power. Both pictures
show a shift from the yellow spectral regime for the transversally
low excited cavity modes located near the trap center to the green
for transversally higher excited modes appearing at the outer trap
regions. In the lower image, a bright spot is visible in the center
with a measured FWHM diameter of $(14\pm2)\,\text{\textmu m}$. Within
the quoted uncertainties, this agrees with the expected diameter of
the $\text{TEM}_{00}$ mode, $d=2\sqrt{\hbar\ln2/m_{\text{eff}}\Omega}\simeq12.2\,\text{\textmu m}$,
yielding clear evidence for a macroscopic population of the transversal
ground mode. Further, we observe that not only the height of the condensate
peak increases for higher photon numbers, but also its width \cite{Klaers:2010}.
This mode diameter increase suggests a weak repulsive self-interaction
mediated by the dye solution. The origin of this is most likely thermal
lensing, but in principle it could also be due to a microscopic Kerr-nonlinearity
in the dye medium. Both effects contribute to the nonlinear index
of refraction of eqn \eqref{eq:E_int}, if they are treated on a mean-field
level. By comparing the observed increase of the mode diameter with
numerical solutions of the two-dimensional Gross-Pitaevski equation,
we estimate a dimensionless interaction parameter \cite{Hadzibabic:2009p2114}
of $\tilde{g}=(7\pm3)\times10^{-4}$. This is significantly smaller
than the values in the range $10^{-2}\ldots10^{-1}$ reported for
two-dimensional atomic gas experiments \cite{Hadzibabic:2006,Clade:PRL2009},
and also below the regime in which Kosterlitz-Thouless physics can
be expected to be relevant \cite{Hadzibabic:2009p2114}. An indication
for the latter would be the loss of long-range order. Experimentally,
when directing the condensate peak through a sheering interferometer
we have not seen any signatures for such a spatially varying phase.
Thus, the observed condensation does not seem to deviate from the
BEC scenario of an ideal or weakly interacting Bose gas.

\begin{figure}
\begin{centering}
\vspace{0mm}

\par\end{centering}

\begin{centering}
\includegraphics[width=0.85\columnwidth]{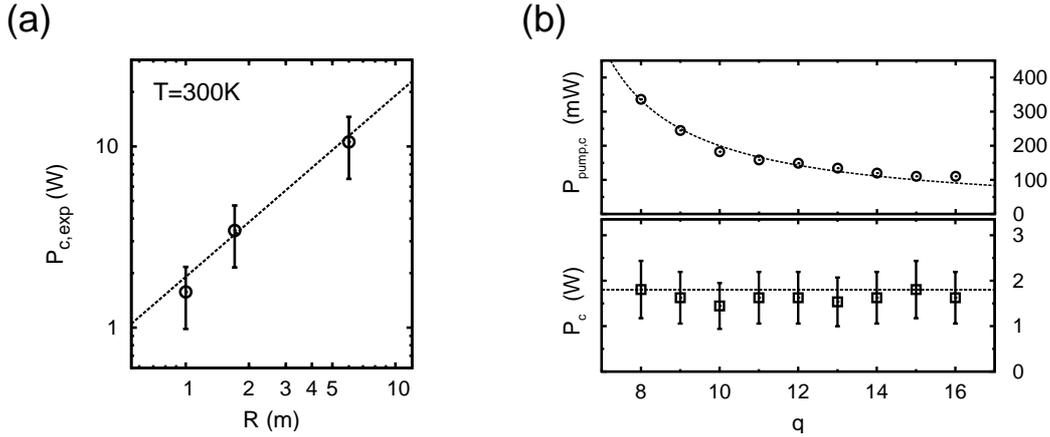}
\par\end{centering}

\begin{centering}
\vspace{0mm}
\caption{\label{fig:criticality}(a) Critical circulating intracavity power
as a function of the effective mirror curvature R. (Rhodamine 6G in
methanol, concentration $1.5\times10^{-3}\,\text{Mol/l}$, cavity
order $q=7$, pulse duration $0.5\,\text{\textmu s}$) (b) Pump power
at the phase transition (top) and critical intracavity power (bottom),
respectively, as a function of the cavity order q. (Rhodamine 6G in
methanol, concentration $1.5\times10^{-3}\,\text{Mol/l}$, mirror
curvatures $R_{1}=R_{2}=1\,\text{m}$, pulse duration $0.5\,\text{\textmu s}$)}

\par\end{centering}

\centering{}\vspace{3mm}

\end{figure}

In further experiments, we have investigated the influence of the
resonator geometry on the observed critical particle number. For that
we measured the critical circulating power for various mirror radii
and different mirror spacings. From eqn \eqref{eq:Nc} we expect that
the critical power $P_{\text{c}}=N_{\text{c}}\hbar\omega^{2}/2\pi q=(\pi^{2}/12)(k_{\text{B}}T)^{2}(\omega/\hbar c)R$
should increase linearly with the mirror radius $R$ and have no dependence
on the cavity order $q$ (at least as long as the two-dimensionality
of photon gas holds). Figure \ref{fig:criticality}a and the lower
graph of Fig. \ref{fig:criticality}b give the corresponding experimental
results, which confirm the theoretical predictions, both the scaling
and the absolute values.%
The upper graph of Fig. \ref{fig:criticality}b shows the pump power
required to achieve criticality as function of the longitudinal mode
number $q$. Here, a decrease is observed for larger mirror spacings.
This becomes clear by noting that the pump power absorption, which
is of order 1\% of the incident light, increases with the thickness
of the dye film. Therefore, less pump power is necessary to reach
a given circulating power. The experimental data can be well modeled
by assuming that the absorbed pump power at criticality remains constant
when increasing $q$ (solid line). Note that a quite contrary scaling
was observed in early work on 'thresholdless' microlasers, for which
an increase of the pump threshold with mirror spacing has been reported
\cite{DeMartini:1988p632,Yokoyama:1992p2123}. 

\begin{figure}
\begin{centering}
\vspace{0mm}

\par\end{centering}

\begin{centering}
\includegraphics[width=0.45\columnwidth]{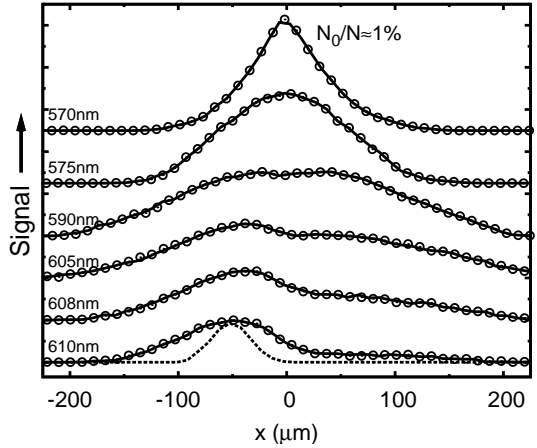}
\par\end{centering}

\begin{centering}
\vspace{0mm}
\caption{\label{fig:spatial_relaxation}Intensity distribution of the photon
gas along an axis intersecting the trap center for different cut-off
wavelengths $\lambda_{c}$ (indicated on the left hand side). The
pump beam (dashed line) is located outside the trap center and its
position as well as its power are kept fixed during the course of
the measurement. The top curve shows a photon gas at criticality with
a ground state population of $N_{0}/N\lesssim1\text{\%}$ (Rhodamine
6G in methanol, concentration $1.5\times10^{-3}\,\text{Mol/l}$, cavity
order $q=7$, mirror curvatures $R_{1}=R_{2}=1\,\text{m}$, pulse
duration $0.5\,\text{\textmu s}$)}

\par\end{centering}

\centering{}\vspace{3mm}

\end{figure}

As already discussed, for lower intracavity powers the thermalization
process is accompanied by a spatial redistribution of photons towards
the trap center. This effect is also observed for higher pump powers,
and we expect it to provide a critical photon density in the center
of the cavity even for a displaced pumping spot. We note that this
effect has been observed in polariton BEC experiments \cite{Balili:2007p1342},
but is unobserved in lasers. For a corresponding measurement, the
pump beam was displaced $50\,\text{\textmu m}$ away from the trap
center. A set of spatial intensity profiles for different values of
the cavity cutoff wavelength is shown in Fig. \ref{fig:spatial_relaxation}.%
By a tuning of the cavity cutoff one can vary the reabsorption probability,
which alters the degree of thermalization. The lowest shown profile
displays results recorded with a cutoff wavelength near $610\,\text{\textmu m}$.
The low dye reabsorption in this wavelength range inhibits an efficient
thermalization, and spatial accumulation at the trap center. Thus,
the maximum of the fluorescence remains at the location of the pump
spot (shown by a dashed line). If the cutoff wavelength is decreased,
the reabsorption probability increases and the intensity profiles
get more symmetrically distributed around the trap center. The top
profile shows data recorded for a cutoff near $570\,\text{nm}$. Here,
the photon gas here seems fully thermalized. The spatial intensity
profile shows a cusp at the trap center, which indicates criticality
(condensate fraction $\lesssim1\text{\%}$). As the pump beam intensity
at the trap center is almost negligible, we conclude that the critical
photon density is not directly generated by pumping, but is a consequence
of the spatial relaxation towards the cavity ground state connected
to the thermalization.

\section{Conclusions}

We have investigated thermodynamic properties of a two-dimensional
photon gas in a dye-filled optical microcavity. A fluorescence induced
thermalization process leads to a thermal photon gas at room temperature
with freely adjustable chemical potential. Evidence for a Bose-Einstein
condensation of photons was obtained from Bose-Einstein distributed
photon energies including a massively populated ground state mode,
the phase transition occurring at the expected critical photon number
and exhibiting the predicted dependence on the cavity geometry, and
a spatial relaxation process leading to a condensation even for a
displaced pump spot. 

An interesting question is, in what respect does photon Bose-Einstein
condensation relate to lasing. In general, the main borderline between
a laser and a photonic Bose-Einstein condensate is that lasing is
a non-equilibrium phenomenon, while Bose-Einstein condensation occurs
in thermal equilibrium. Thermalization of the photon gas above a cavity
low-frequency cutoff can be achieved in media fulfilling the Kennard-Stepanov
relation, when recapturing (spontaneous) emission in one of the many
transversal cavity modes, and tuning to a regime with strong reabsorption.
In such a system, the condition for the emergence of a macroscopically
occupied photon mode, which for a laser usually is given by a (small-signal)
gain condition in an inverted medium, is replaced by a Bose-Einstein
criticality condition, as given in eqn \eqref{eq:Nc} for the case
of a harmonically trapped two-dimensional photon gas. In essence,
this is equivalent to the condition of the phase space density exceeding
a value near unity. The thermalization process will select the cavity
ground state to have the biggest statistical weight, which even becomes
macroscopic when condensation sets in.

For the future, we plan to investigate condensate properties in more
detail. For example, an intriguing question is whether the observed
self-interaction is sufficient to cause superfluidity. Also, the coherence
properties of the photon condensate should be further explored. Initial
interferometric measurements have already delivered a lower bound
for the first order coherence length in the centimeter regime \cite{Klaers:2011,Klaers:Thesis2011}.
It will be important to also investigate the second-order coherence
of the photon Bose-Einstein condensate. This is of particular interest
because of the unusually large intensity fluctuations that are expected
for grandcanonical Bose-Einstein condensates.

\bibliography{references}

\end{document}